\documentclass{aa}  
\usepackage{graphicx}
\usepackage{natbib}
\usepackage{longtable,lscape}
\usepackage{rotating}

\usepackage{txfonts}
\newcommand{\msun}{\mbox{$\rm M_\odot\,$}}
\newcommand{\Msun}{\mbox{$\rm M_\odot\,$}}

\newcommand{\mum}{\mbox{$\rm \mu m\,$}}

\def\deg{\hbox{$^\circ$}}
\def\arcmin{\hbox{$^\prime$}}
\def\arcsec{\hbox{$^{\prime\prime}$}}

\begin{document}
\title{H$_2$ flows in the Corona Australis cloud and their driving sources} 
\subtitle{}
   \author{M. S. N. Kumar\inst{1}, Saurabh Sharma\inst{2,5}, C. J . Davis\inst{3,4}, J. Borissova, \inst{2}
          \and J. M. C. Grave \inst{1}
		}

   \offprints{M. S. N. Kumar; email:nanda@astro.up.pt}

   \institute{Centro de Astrof\'{i}sica da Universidade do Porto, Rua
     das Estrelas, 4150-762 Porto, Portugal 
     \and Departamento de F\'isica y Astronom\'ia, 
        Facultad de Ciencias, Universidad de
        Valpara\'{\i}so,Ave. Gran Breta\~na 1111, Playa Ancha, Casilla
        53, Valpara\'iso, Chile
     \and Joint Astronomy Center, 660 N. A'ohoku Pl, Hilo, Hawaii 96720, USA
     \and NASA Headquarters, Science Mission Directorate, 300 E St SW,
          Washington, DC 20546, USA
     \and Aryabhatta Research Institute of Observational Sciences, Nainital 263129, India\\
     \email{nanda@astro.up.pt}} \authorrunning{Kumar et al.}
   \titlerunning{H$_2$ flows in the Corona Australis and their driving sources}
 %  \date{Received September 15, 1996; accepted March 16, 1997}
\date{\today}
% \abstract{}{2}{3}{4}{} 
% 5 {} token are mandatory
 
  \abstract
% context heading (optional)
 {} 
% aims heading (mandatory)
{We uncover the  H$_2$ flows in the Corona Australis
  molecular cloud and in particular identify the flows from the
  Coronet cluster.}
% methods heading (mandatory)
{A deep, near-infrared H$_2$ v=1--0 S(1), 2.122$\mum$-line,
  narrow-band imaging survey of the R CrA cloud core was carried
  out. The nature of all candidate-driving sources in the region was
  evaluated using data available from the literature and also by
  fitting the spectral energy distributions (SED) of each source
  either with an extincted photosphere or YSO model. Archival {\em
    Spitzer}-IRAC and MIPS data was used to obtain photometry, which
  was combined with USNO, 2MASS catalogs and millimeter photometry
  from the literature, to build the SEDs. We identify the best
  candidate-driving source for each outflow by comparing the flow
  properties, available proper motions, and the known/estimated
  properties of the driving sources. We also adopted the thumbrule of
  outflow power as proportional to source luminosity and inversely
  proportional to the source age to reach a consensus.}
% results heading (mandatory)
{Continuum-subtracted, narrow-band images reveal several new Molecular
  Hydrogen emission-line Objects (MHOs). Together with previously
  known MHOs and Herbig-Haro objects we catalog at least 14
  individual flow components of which 11 appear to be driven by the R
  CrA aggregate members. The flows originating in the Coronet
  cluster have lengths of $\sim$0.1-0.2\,pc.  Eight out of nine
  submillimeter cores mapped in the Coronet cluster region display
  embedded stars driving an outflow component. Roughly 80\%
  of the youngest objects in the Coronet are associated with
  outflows. The MHO flows to the west of the Coronet display lobes
  moving to the west and vice-versa, resulting in nondetections of
  the counter lobe in our deep imaging. We speculate that these
  counterflows may be experiencing a stunting effect in penetrating
  the dense central core.}
% conclusions heading (optional), leave it empty
{ Although this work has reduced the ambiguities for many flows in
    the Coronet region, one of the brightest H$_2$ feature (MHO2014)
    and a few fainter features in the region remain unassociated with
    a clear driving source. The flows from Coronet, therefore, continue to
    be interesting targets for future studies.}

   \maketitle
%
%________________________________________________________________

\section{Introduction}
The Corona Australis molecular cloud \citep[see the review
  by][]{neu08}, located in the southern hemisphere of the sky, is one
of the nearest (130-170pc; recommended value is 130pc) and best
studied star-forming regions with intense star formation activity. The
1.2mm dust continuum emission, mapped by \citet{chini03} using the
SIMBA array on the SEST telescope, provides a good representation of
the dense regions of this cloud where most of the active star
formation takes place. The densest region of this cloud roughly
coincides with the Herbig Ae/Be star R Corona Austrinae (RCrA),
surrounded by a cluster of young stars, well-known as the Coronet
cluster \citep{taylor84}.  The Coronet cluster and associated
molecular core has been the target of several studies at different
wavelengths, from X-rays \citep[for a complete list, see the review
  by][]{neu08} through the optical and infrared \citep{wilking97} to
the submillimeter (submm) \citep{nutter05}.

Outflows and jets are ubiquitous phenomena associated with star
formation activity. Bipolar outflows are unambiguously identified by
mapping the two lobes using molecular line transitions from species
such as CO or SiO at (sub)mm wavelengths.  Observations of optical
Herbig-Haro (HH) objects and/or near-infrared (near-IR) H$_2$ v=1--0
S(1) line emission features often complement these radio data; indeed,
they can be more successful in tracing the flow components,
particularly in crowded regions, because of the higher spatial
resolution typically available at these shorter wavelengths.

Molecular line maps of the Coronet region show a prominent east-west
bipolar outflow and a weak north-south flow
\citep{anderson97,groppi04,groppi07}. The east-west flow is thought to
be driven by sources around IRS7, which is located roughly
15-20\arcsec\ southeast of RCrA.  The weak north-south flow is driven
by the IRS1 source.  This outflow is associated with HH~100, and its
driving source is often referred to as HH100-IR.  In contrast, at
least 20 HH objects are known to be located in the close vicinity of
the Coronet cluster \citep[see][and references therein]{wang04}.
These data suggest the presence of several flow components. The
east-west molecular outflow has been mapped in a number of molecular
lines (e.g.  CO, HCO+, and SiO), and these observations likewise infer the
presence of more than one outflow component
\citep{anderson97,groppi04}. However, decomposition of these flow
components has not been possible using the lower spatial resolution
radio data.

The optical observations of \citet{wang04} and, more recently, the
near-infrared H$_2$ images of \citet{cog06} and \citet{pet11} provide
a far superior view of outflow activity in the region.  Even so, the
concentration of candidate driving sources within a small region in
the Coronet cluster, together with the large number of flow
components, has made it difficult to unambiguously associate the flows
with their driving sources.  The rather complete census of the
embedded population of young stars in the region uncovered by the new
X-ray \citep{for07} and {\em Spitzer Space Telescope} observations
\citep{sic11,pet11} has prompted a fresh look at this region.

In an attempt to effectively trace the outflow activity in this region, we
have conducted a deep H$_2$ v=1--0 S(1) 2.12\mum narrow-band imaging
survey of the northern part of the Corona Australis cloud. Many new
Molecular Hydrogen emission-line Objects  \citep[MHOs][]{dav10} are
discovered from the new data, increasing the number of known flow
components. In this work, we present and organize all the new and
previously known outflow components. 

 For this purpose we use the following strategy. It has been
 suggested that the outflow power (and therefore MHO activity) is
 directly proportional to the luminosity and inversely to the
 evolutionary stage of their driving sources
 \citep[e.g][]{bontemps96}.  First, by using the available information
 in the literature and then by modeling the spectral energy
 distributions (SED) of candidate driving sources in the region, we
 obtain a handle on the luminosity (mass) and evolutionary stage
 (age).  Combining this information with the known properties of the
 outflows, as stated above, we then disentangle the individual
 flows and attribute the best driving sources to each flow. We
 also use the information on the proper motions of many knots in this
 region presented by \citet{pet11} to arrive at the best-case
 scenario.

\section{Observations and data analysis}

\subsection{Narrow-band H$_2$ imaging}

Near-infrared observations of the Corona Australis molecular cloud
were carried out using the 4m Vitor Blanco Telescope at the
Cerro-Tololo Inter-American Observatory (CTIO) on the nights of
11 to 13 July 2009.  Imaging observations in the H$_2$ 2.12\mum
narrow-band filter and in the 2.2\mum broad-band K filter were
obtained. These observations cover the northern part of dense
  gas regions traced by the 1.2mm observations of \citet{chini03}.
The near-infrared wide-field camera ISPI, used to obtain the
observational data, has a 10\arcmin\ field of view and a plate scale
of 0.3\arcsec\ per pixel. A nine-point jitter pattern was used to
obtain the observations, and the jitter sequence was repeated three times
at each pointing of the telescope in each filter. An exposure of 60
sec per jitter position was used for the H$_2$ observations, resulting
in a total integration time of 27 mins. For the K-band observations,
the exposure time per jitter position was kept at 5 sec, resulting in
an integration time of 2 min.

Each set of jittered images were median combined to obtain a skyframe
that was subtracted from each individual image. The sky-subtracted
images were subsequently flattened by dividing by a smooth
surface-fit obtained from the image.  The individual exposures were
co-added and mosaicked together using KAPPA and CCDPACK routines in
the Starlink\footnote{http://starlink.jach.hawaii.edu} suite of
software.

\subsection{ {\em Spitzer} photometry}

Archival {\em Spitzer} imaging data were used to obtain photometry of
the sources in the Coronet region, which in turn were used for SED
modeling.  The data used correspond to the astronomical observational
requests (AOR's) 3650816 and 17672960 (PI: G. Fazio) and 27041280 (PI:
L. Allen). The BCD data in each of the four IRAC channels were
corrected using artifact mitigation software provided by the Spitzer
Science Center (SSC), and were then mosaicked together to obtain
images corresponding to the region covered by our H$_2$ narrow-band
observations. We used the {\em APPHOT} package in {\em IRAF} to obtain
aperture photometry using an aperture of 2.4\arcsec and a sky annulus
of width 12\arcsec\ . The aperture corrections used for the Ch1-4
  IRAC bands are 1.213, 1.234, 1.379, and 1.584.

MIPS 24\mum\ data of the same region were obtained from AOR 3664640
(PI: G. Fazio). The BCD data were mosaicked using standard templates
available on MOPEX software. Aperture photometry was performed by
using the MOPEX/APEX single frame template.  Aperture size of
  6\arcsec, sky annuli of width 7\arcsec and an aperture correction
  factor of 2.05 were used.

\section{Spectral energy distributions and modeling}

 As mentioned in the last paragraph of Sec.1, we will first obtain the
 information on luminosities (masses) and evolutionary states (ages)
 of candidate driving sources in the Coronet region. For this purpose,
 we first searched the literature to obtain the information derived
 from standard methods. When such information was not
 available, particularly for the candidate sources that are
 well-detected in the {\em Spitzer} observations, we analyzed their
 spectral energy distribution by modeling. By SED fitting, we first
 eliminated sources that display redenned photospheres and then used the
 SED modeling to make a relative comparision of YSOs based on the
 fitted parameters (class, luminosity, and accretion rates) to select
 the most probable driving source for each identified outflow.

To this end we first built the SEDs of sources that appear to be
associated with outflows and model these with a popular SED fitting
tool \citep{rob07}. Not all SEDs are uniformly populated with the same
photometric data because many of the millimeter sources are too bright
(saturated) in the infrared bands. Some sources are resolved by
optical and/or near-IR data and unresolved in the millimeter
bands. Also, while we modeled all candidate driving sources in the
mapped region, our focus is on a selected group of sources, which
appear to drive outflows (see Table.\,2). These sources are prominently
infrared-millimeter visible and were all modeled with 4-6 ``data
points'' and with 1-6 upper or lower limits.  The photometric data,
apertures used, and the source of the data are listed in Table.\,
1. Only the IRAC and MIPS photometry are independently derived by us,
and the remaining photometric fluxes and errors are directly obtained
from the references listed in Table.\,1. The apertures listed in
Table\, 1 are those used for SED fitting and not necessarily the one
with which photometry was obtained. This is because the SED fitter
requires that the modeled source should not be larger than the
aperture used for fitting.  If a source is saturated in any band, the
flux is assumed to be a lower limit and if the source is too faint or
larger than the aperture used, then the flux is assumed to be an upper
limit. The 2MASS magnitudes are used as an upper limit if either the
confusion flag was not ``0'' (implying confusion or extended source)
or if the photometric quality flag was not the best (anything other
than A). Ten percent errors are assumed for fluxes for the shorter wavelength
bands (up to 8\mum\ ) and 20\% errors were used for all other longer
wavelength data. In most cases, the photometric accuracy is better
than the assumed error.

\begin{table}
 \centering
 \caption{Data used for SED fitting}
  \label{tab:aperture}
\begin{tabular}{lccccc}

  \hline\hline\\
Band & Aperture & Reference \\

\hline
USNO B, R, I & 3\arcsec & \\
2MASS J, H, K & 2.4\arcsec & \\
{\em Spitzer} IRAC & 3\arcsec & this work \\
{\em Spitzer} MIPS 24\mum & 6\arcsec & this work \\
{\em Spitzer} MIPS 70\mum & 9\arcsec & this work \\
450\mum SCUBA & 8\arcsec & \citet{nutter05}\\
850\mum SCUBA & 14\arcsec & \citet{nutter05}\\
1.2mm SIMBA & 24\arcsec & \citet{chini03}\\
1.1mm SMA & 3\arcsec & \citet{groppi07}\\  

\hline
\end{tabular} 
\end{table}

The resulting SED data were fed into the online SED fitting tool
described by \citet{rob07}. The distance and extinction values were
allowed to vary in the range 130--170\,pc and 5--45\,A$_v$
respectively. The SED fitting tool attempts to fit the observed
fluxes, either with an extincted photosphere or an young stellar
object (YSO).  Models are constrained by the goodness of fit,
${\chi}^2$. The YSO models used were obtained from a grid of 200,000
models, as described by \citet{rob06}.  Following the recommendations
by \citet{rob07}, only models that satisfy the criteria
${{\chi}^2}_{best} - {\chi}^2$$\leq$3 per data point, where
${{\chi}^2}_{best}$ is for the best-fit model, are used in estimating
representative values of some physical parameters.  These are
essentially the group of models displayed in Fig.~4.  The results
  using this group of models are listed in Table.\,3. The
  minimum, maximum, and weighted means (weights being the inverse of
  $\chi^2$) are listed. It should be noted that the weighted means
  represent the parameter value distribution better than the
  difference in the maximum and minimum values. This is because even a
  single outlier model or parameter within the chosen ${\chi}^2$
  criteria can produce a minimum and a maximum that are not representative
  of the overall distribution. The minimum and maximum values
  nevertheless represent the extreme values for a given parameter
  from the group of models shown in Fig.\,4. The evolutionary stage
of Class 0/I quoted in Table\,3 represents a fitted source age in the
range 1--10$\times$10$^4$yrs.

 In what follows we present the results from this procedure only for a
 few selected sources, for which the data available in the literature
 were not decisive in associating with detected MHOs. SEDs that
 were fitted with extincted photospheres are also not shown, except for
 one example, IRS11 (see Fig.~4), as they were ruled out of
 qualifying as driving sources.

\section{The molecular yydrogen emission-line objects (MHOs)}

Figure\,1 (online only) is a color composite generated by using the
{\em Spitzer} IRAC-4.5\mum\, image as red, the H$_2$ narrow-band image
as green and the SII narrow-band image (mosaicked using archival ESO
data from \citet{wang04} as blue. The SII data do not cover the lower
portion of the image. The scaling of the individual images is
adjusted to effectively display the embedded sources and the outflow
components simulatneously. Since [SII] and H$_2$ trace different gas
excitation conditions, they are often hugely complementary when one is
trying to trace large sections of a protostellar outflow and/or
connect outflow features with distant driving sources.

The continuum subtracted H$_2$ image of the mapped region is displayed
in Figure\,2 where five groups of MHOs are marked by ellipses. The
majority of the MHOs are contained within the large central
ellipse. In Table\,2 we list the positions of the MHOs and the
identified driving sources, along with the measured flow lengths in
arcseconds. Of these, the newly detected MHOs 2005, 2007, 2014, and 2017
are located away from the Coronet cluster and the dense core
associated with it. The MHO features in the central region,
corresponding to the Coronet cluster, are shown in Figure.\,3.

 The Coronet region is composed of many previously known infrared
 sources plus new infrared counterparts for the millimeter sources
 \citep{for07,sic11,pet11}. Notably, SMM4, SMA1, and SMM2 have been
 recently detected by {\em Spitzer} images. This demonstrates that
 these submm cores are protostellar in nature. Furthermore, the SMA
 observations by \citet{groppi07} and \citet{pet11} that provide
 millimeter fluxes are very useful in identifying the nature of the
 youngest protostars in the Coronet region. To disentangle each MHO
 flow and associate it with the best candidate driving source, we used
 the results of our SED modeling of all {\em Spitzer-IRAC} detected
 point sources (shown using circular symbols in Figure\,3) and
 utilized the proper motion information provided by
 \citet{pet11}. Together, these data have significantly improved the
 association of driving sources to the detected MHOs in the Coronet
 region. In Figure\,4 we show the observed SED data and the fitted
 models for candidate driving sources. The results of our SED modeling
 are presented in Table\,3.  In the following, we first discuss MHO
 features that have driving sources reasonably well identified.\\

\subsection{ MHOs with identified driving sources}

{\it MHO 2000/HH 99 -- SMA2:} MHO2000 is the brightest,
well-collimated bow shock feature to the north of Coronet, associated
with HH99. The suggestions of \citet{wang04} that it is driven by
IRS6, and of \citet{dav99} that it is driven by IRS9 were ruled
out by \citet{cog06} based on the evolved nature of IRS6 and also on the
position angle of proper motion vectors. Recently, \citet{pet11}
suggest that it is driven by IRS7 or CXO34. The position angle of
the proper motion vector for MHO2000 was found to be 39$\pm$6$\deg$ by
\citet{cog06} and 43$\pm$6$\deg$ by \citet{pet11}. These position
angles of the proper motion vectors clearly point towards the sources
SMA2, IRS7, and CXO34.

 Since MHO2000 is a bright and well-collimated feature, we assume that
 it has to be driven by a Class0/I type object in the region. SMA2 is
 bright at millimeter wavelengths, very bright (nearly saturated) at
 24$\mum$, and not detected at 8$\mum$ IRAC images, therefore, has
 insufficient data points to be modeled. Nevertheless, the above
 characteristics are strong enough to classify it as a Class 0
 protostar \citep[see also][]{groppi07,pet11}. In contrast, IRS7A
 display characteristics of being more evolved than SMA2 since it is
 bright at 8$\mu$m, not detected at (sub)millimeter, and also modeled
 to have a very low disk mass (0.004\,\Msun). Similarly, CXO34 is
 modeled as a substellar object with very low disk mass as
 well. In fact, using the SMA nondetections (both IRS7A and CXO34),
 \citet{pet11} suggest an upper limit of only a few Jupiter masses for
 the disk in these sources. On the contrary, they estimate a disk mass
 of 0.063\,\Msun\, for SMA2, higher by a factor of ten. Therefore, we
 conclude that SMA2 drives the MHO2000 flow, which is a bright,
 well-collimated bowshock.\\

{\it MHO 2003,2008 (2007?) --IRS1/HH100IR:} The source IRS1, also
known as HH 100IR is known to drive the Herbig-Haro objects HH 100 and
HH 97 \citep{wang04}. \citet{cog06} discovered MHO 2003 (an arc and an
elongated knot) associated with IRS1. The deeper images presented here
and by \citet{pet11} reveal a new knot named as MHO2008, which is
probably the tip of the flow associated with IRS1. While MHO2003 and
MHO2008 are certainly driven by IRS1, it is likely that this flow is
much longer that it terminates in MHO2007 (see the description of this
MHO in the next subsection). MHO2003 appears to have a wide opening
angle, MHO2009 from IRS2 and the confused flows from RCrA being the
only two other cases in the region with large opening
angles. \citet{nisini05} estimate a spectral type of K-M for the star,
however, they point out that IRS1 and IRS2 are the only two sources
with a bolometric luminosity that is 60-80\% higher than the stellar
luminosities. Also, using Brackett line characteristics, they suggest
it probably originates in an expanding stellar wind rather than in
accreting regions. The arc-shaped MHO feature 2003 may therefore be
due to fluorescence rather than due to a large opening angle of the
flow. \\

{\it MHO 2005 C, D, E--S CrA:} This group of MHO knots represents the
H$_2$ counterpart of the well-known optical features HH 82A, HH 82B,
and HH 729A,B,C \citep{wang04}, which are located near the bright
T-Tauri star S CrA. This star coincides with the millimeter peak MMS1
described by \citet{chini03}. Previous studies have suggested that
these HH objects are driven by S CrA \citep{rg88,wang04}, and so are
these MHO knots. The knots MHO2005A,B \citep{pet11} are not associated
with S CrA, and is probably driven by IRS5a/b as discussed in
Sect\,4.2.\\

{\it MHO 2006/HH 730--IRS2:} This source, first identified by
  \citet{wilking97}, is classified as a young star with a mass of
  1.4\msun\, by \citet{nisini05}. The recently discovered MHO 2006
  feature \citep{pet11} is clearly associated with this source and has
  a wide opening angle, most likely due to reasons like those discussed
  above with respect to IRS1. The HH objects HH 730A, HH 730B, and HH
  730C, which terminate in a well-defined bowshock, clearly align
  with this source. We therefore associate HH 730 and the newly
  detected MHO 2006 object with a single flow driven by IRS2.\\

{\it MHO 2011 A, B \& 2012 -- SMM2:} The newly discovered knots 2011 A,
B, and the 2012 are located to the west of the Coronet cluster. It
comprises a number of knots and arcs that probably form part of at
least two bow shocks. This flow is well aligned with the embedded IR
source associated with the submm peak SMM2. The SED modeling presented
here indicates that the {\em Spitzer} detected infrared counterpart to
SMM2 is a young low mass star of Class 0/I type (see Table.\,3).\\

{\it MHO 2013/2015/HH733 -- TCrA:} The faint bowshock MHO 2013
\citep{dav99} by the side of MHO2000 is the H$_2$ counterpart of the
Herbig-Haro object HH 733.  \citet{wang04} suggest that this flow is
driven by TCrA. The newly discovered MHO 2015 is a clear bow-shock
feature, lying to the south of TCrA, and it marks the southern lobe of the
bipolar outflow originating from TCrA. This is probably the only
unambiguously detected bipolar outflow traced by two complementing
bow-shock features in the entire Coronet region. \citet{mw09} adopt a
spectral type of F0 for TCrA.\\

{\it MHO2008D -- Star1-100:} The star identified as 1-100HH by
\citet{wang04} is associated with a newly discovered faint H$_2$ knot
that is labeled as MHO2008D by \citep{pet11}. This is very likely
an independent flow driven by this star.\\

{\it MHO 2016:} This MHO feature has two conical nebulae A and B
centered on TY CrA and HD176386, respectively. MHO 2012A is an extended
conical nebula, centered on the H$_{\alpha}$ emission star TY CrA. MHO
2012B is a relatively faint, localized emission on the binary
pre-main-sequence star HD176386.  \\

\subsection{ MHOs with ambiguous driving source associations}

{\it MHO 2001 -- IRS6b, CXO34, IRS5N:} MHO 2001 is one of the
brightest features to the east of Coronet cluster, well immersed in
the reflection nebula. There is much confusion in this area. Since it
is one of the brightest MHO features, it should be driven by a
relatively young object of Class 0/I type. However, it appears to be
close to most of the candidate driving sources (unlike MHO2000 which
is farther away). \citet{pet11} quote CXO34 as a possible candidate
driving this flow. However, the same authors quote a position angle of
the proper motion vector at 81$\deg$ for the bright knot MHO2001A,
which when traced backwards, points more directly to IRS6 or roughly
towards IRS5N (considering an average of the PA from all the knots
associated with MHO2011). IRS6 is known to be a binary source with one
young and one old component \citep{nisini05}. Following the
proper-motion vectors, a likely scenario is that the younger object of
this binary component drives MHO2001. Alternatively, MHO2001 could be
driven by SMM4/IRS5N, which is a millimeter peak associated with a
faint IR source, that is modeled as a very young substellar
object. We also point out the low flux density of the SMA peak associated with
IRS5N in comparison to SMA1 or SMA2 \citet{pet11}. We favor the
scenario of MHO2001 driven by IRS6 rather than IRS5N because of the
likely stunting effect discussed in Sect.5 \\

{\it MHO 2002, 2004 -- RCrA :} RCrA is the principal component of the
Coronet cluster. We associate the previously detected
bow-shaped object MHO 2002 with RCrA \citep[labeled "E" by][]{cog06} and
possibly a knot from within the group labeled MHO 2004.  Together,
these would delineate an east-west bipolar outflow.  These knots
appear as a bowshock from a relatively wide-angled wind.  The
morphology of these MHO features suggests a wide-angled, poorly
collimated wind rather than a collimated outflow.  RCrA is a bright
source that is saturated in all of the {\em Spitzer} bands. These
properties conform to the nature of RCrA as a Herbig Ae/Be star
with an estimated mass of 5-10\msun \citep{tuthill01}.  Recently,
\citet{kraus09} have used the {\em Very Large Array (VLA)}
Interferometer to study this source and find a disk that is oriented
north-south, suggesting an east-west outflow. The proposed MHO
associations with RCrA presented above are in good agreement with this
scenario.\\

{\it MHO 2004/HH 104 -- IRS6b or RCrA:} We associate HH 104 and MHO
2004 with a faint, jet-like feature that is located close to the
source IRS6.  \citet{cog06} suggest that IRS6 is too evolved to be
driving a powerful outflow. In their high-resolution VLT data,
\citet{nisini05} identify IRS6 as a binary system (IRS6a and IRS6b),
and by using photometric data, show that IRS6b has a stronger H-K color
than IRS6a.  Our K band data also partially resolve this binary
source. Therefore, the HH 104/MHO 2004 outflow is probably driven by
the fainter -- though redder -- infrared source, dubbed IRS6b by
\citet{nisini05}. IRS6 is unresolved in the mid-IR and submm
photometric data.

MHO2004 also contains a bright bow-like knot that indicates a proper
motion vector moving to the west; therefore, it could just as well be
due to RCrA flow. Given that RCrA is a Herbig Ae/Be type star, the
outflow from such a star would be composed of strong stellar winds and
contribute to poorly collimated flows composed of many knots and shock
surfaces.\\

{\it MHO 2005 A, B -- IRS5a/b:} The bright infrared source IRS5
\citep{wilking97} is found to be a subarcsecond binary source
\citep{chen93,nisini05}. The main component is labeled IRS5a by
\citet{nisini05} and classified as a Class I source with low accretion
activity. The faint, though well-defined, elongated MHO feature MHO
2016 (Fig.\,3) that terminates in a small bow shock is aligned well
with IRS5a. The Class I nature and the low accretion activity
attributed to this source by \citet{nisini05} is consistent with the
low brightness and morphology of MHO2016. \citet{pet11} associate this
feature with the prominent east-west CO flow.  Although IRS11
\citep{wilking97} is located along the same projected line as MHO
2016, it is unlikely to be the driving source since its SED represents
a reddened photosphere. IRS11 is probably a reddened background star
that is unrelated to the Coronet cluster.\\

{\it MHO 2007 -- IRS1/HH100IR:} MHO 2007 is a group of knots discovered
in the images of \citet{pet11} and our H$_2$ images. Additionally, our
deep H$_2$ data reveal a faint, but clear bow shock feature at the
same location (see Fig.\,1). The bright knots of MHO2007 are the H$_2$
counterparts of the HH objects HH 101A,B,N, and S. \citet{pet11}
associate this group of knots to the red lobe originating in
IRS5N. Also, there is a star embedded in this group of MHO features
/HH objects that was labeled Star30 by \citet{hl85}. \citet{hg87}
argued that Star30 is not the driving source for the HH 101 group of
knots. From our SED fitting we concur with this viewpoint: using the
USNO and 2MASS magnitudes we find that the photometry of star30 is
best described by an extincted photosphere rather than a
YSO. \citet{hl85} and \citet{hg87} suggest that HH 101 must be part of
a long collimated outflow originating in HH 100IR (close to the
Coronet cluster). The faint, but clear, bow-shock feature detected by
our H$_2$ data could indeed be the terminating point of the flow from
HH100IR. In Fig.\,1, the tip of the arrow drawn from IRS1 points to
the center of this bow-shock feature.  Thus, the speculation of
\citet{hg87} and \citet{hl85} can also represent a valid
scenario. This flow from HH 100IR/IRS1 would then measure
$\sim$365\arcsec , corresponding to a single lobe length of
$\sim$0.23\,pc. In such a case, HH 100IR/IRS1 may be driving a parsec
scale outflow.\\

{\it MHO2011 D, E -- SMM1b/SMA1:} The submm source identified by
\citet{nutter05} as SMM1b and resolved as SMA1 by \citet{groppi07}
appears as a point source in the {\em Spitzer-IRAC} images. The
luminosity (see Table.\,4) of this source is relatively higher
compared to other modeled sources in the region. The newly detected
bow-shock features MHO 2011, D \& E are well aligned with this source
in projection. Further, the PA of the proper motion vector estimated
by \citet{pet11} points away from this source.  The source
SMM1c/SMA2 appears to be aligned with MHO 2011, but is a less
probable option because: a) we have already shown that SMA2 very
likely drives MHO2000, b) SMA1 is thought to be relatively older (a
transitional Class0/I object) than SMA2 (Class 0 protostar)
\citep{groppi07,for07}(assuming a similar mass), therefore expected to
be associated with a longer and relatively fainter flow. MHO2011, D,
E therefore are best associated with SMA1. \\

{\it MHO 2014 -- SMM6:} This is a bright extended H$_2$ feature, with
a bow-like component (to the east; lefthand side in Fig.\,2) and a
clumpy component (western half). The bow-like components are
associated with the previously known Herbig-Haro features HH 732 A, B,
and C \citep{wang04}. These HH objects are not identified with any
clear driving source. \citet{pet11} compute the position angle (PA)
for the proper motions to be in the range 31\deg-34\deg$\pm$15\deg and
associate it with the blue lobe originating in IRS5N. They suggest
MHO2007 as the red lobe of this flow. This scenario may appear likely,
because MHO2007, MHO2014 and SMM4/IRS5N align along a straight line
and SMM4 displays a peak at millimeter wavelengths. However, the 1.2mm
flux (and the resulting disk mass) of IRS5N, measured by the same
authors, is weaker than sources such as SMA1 or SMA2. Furthermore,
this source is visible in the infrared bands and is not very
luminous. Our SED modeling suggests that SMM4/IRS5N is a substellar
object with a luminosity $\sim$ 0.5\,L$_{\odot}$. These data together
do not support the scenario of IRS5N driving a powerful, parsec scale
outflow comprising of MHO2014 and MHO2007. It should also be noted
that \citet{pet11} associate SMM4 with IRS5 and state that the new
peak detected using SMA observations is associated with IRS5N, while
SMM4 and the new SMA peak are indeed the same, coinciding with IRS5N.

Here we propose SMM6 (MMS10-11) as an alternative driving source for
MHO2014 \citep{nutter05,chini03}. \citet{chini03} note that there are
no near-IR or FIR sources associated with MMS10-11
peaks. \citep{nutter05} discard the proposition of \citet{chini03}
that MMS6 is a deeply embedded protostar based on their interpretation
of this source as a prestellar object. This interpretation may be
valid, since the {\em Spitzer} MIPS 24\mum or 70\mum images do not
reveal point-like or extended source associated with SMM6. However,
the 160\mum emission in this region is bright and saturated, roughly
following the morphology of the 850\mum emission from
\citet{nutter05}.  Our supposition that SMM6 is the driving source for
MHO2014 requires a PA of 41\deg, which is well within the value and
error bars quoted by \citet{pet11}. Given that this is one of the
brightest MHO in the region conforms to the proposition that it should
be driven by a deeply embedded protostellar object invisible at the
infrared wavelengths.

The western half (righthand side in Fig.\,2) of MHO 2014 is very clumpy
and not likely to be driven by a well-collimated Class0 jet/outflow. The
morphology of this clump and its location in the vicinity of the bright
young stars TY CrA and HD 176386 may imply a different origin. TY CrA
is a well known variable star of Orion type, and HD176386 is a
pre-main-sequence double star coinciding with a submm peak
\citep{chini03}. Lacking proper motion data, we therefore speculate
that this clumpy knot might be due to an unidentified component of the
TY CrA or due to the double components of HD176386. \\

\onlfig{1}{ 
  \begin{sidewaysfigure*}
\vskip 17cm
  \includegraphics[width=24cm]{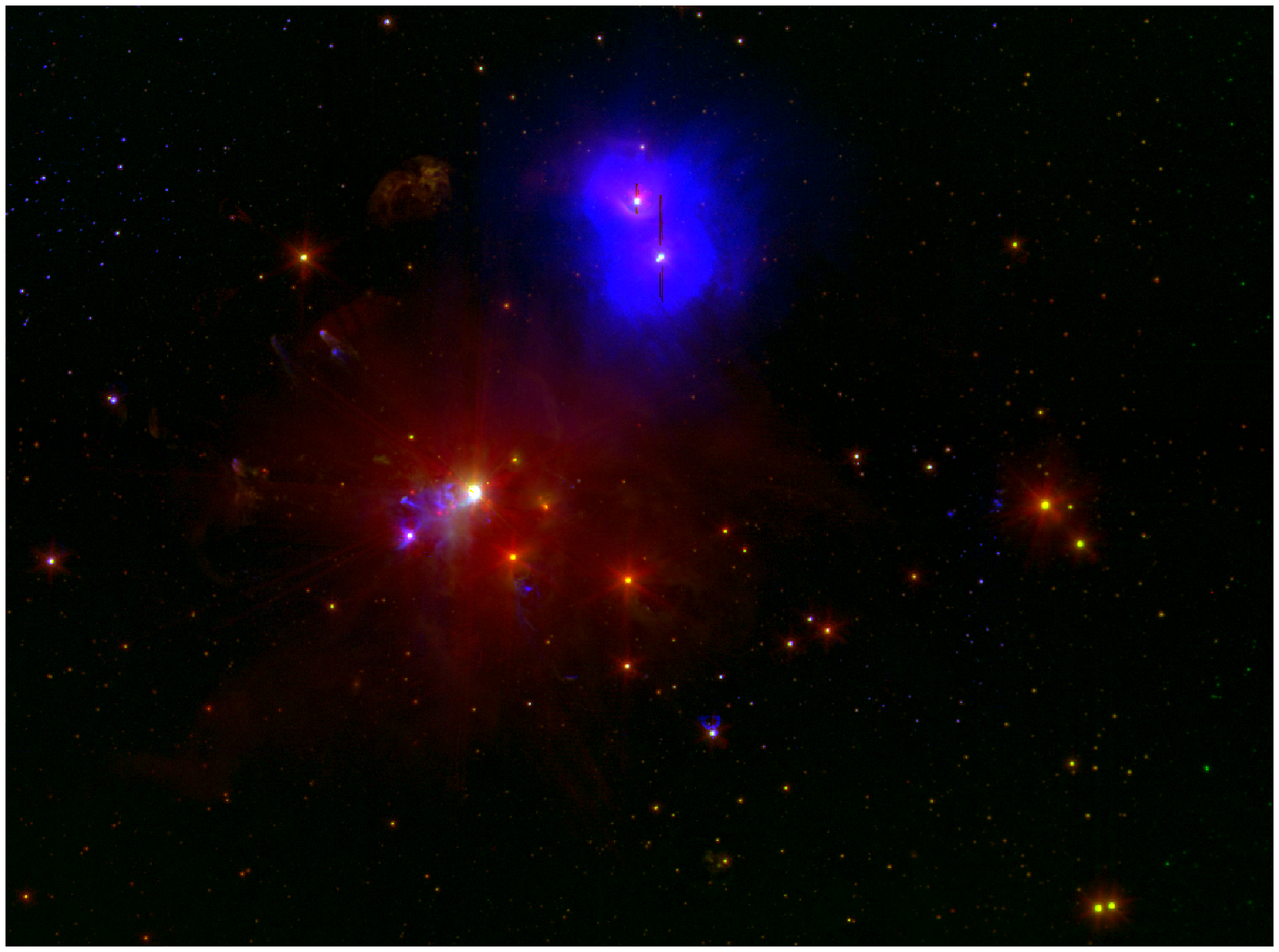}
  \caption{A coloured image of the region studied in this work. The
    colour image was obtained by coding {\em Spitzer} IRAC 4.5\mum\,
    image as red, H$_2$ narrowband image as green and SII image
    \citep{wang04} as blue. The SII data do not cover the lower
    portion of the image. }
  \end{sidewaysfigure*}}

\begin{figure*}

     \includegraphics[width=18cm,angle=0]{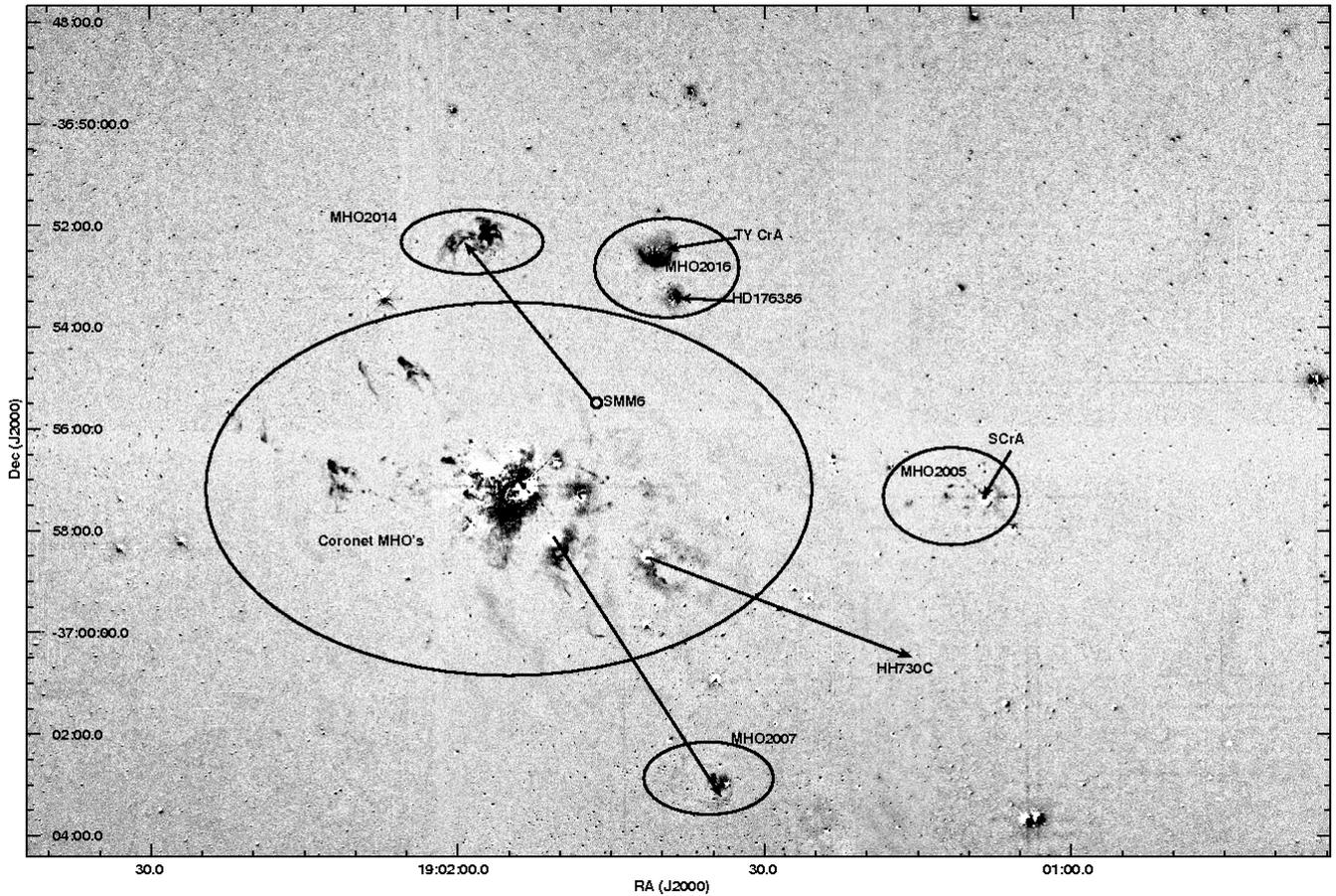}
     \caption{Continuum-subtracted H$_2$ image of the mapped region,
       displayed with linear gray scales. Five groups of H$_2$ line
       emission features are marked by ellipses.}
     \label{fig:2}
\end{figure*}

\begin{sidewaysfigure*}
\vskip 15cm
     \includegraphics[width=24cm,angle=0]{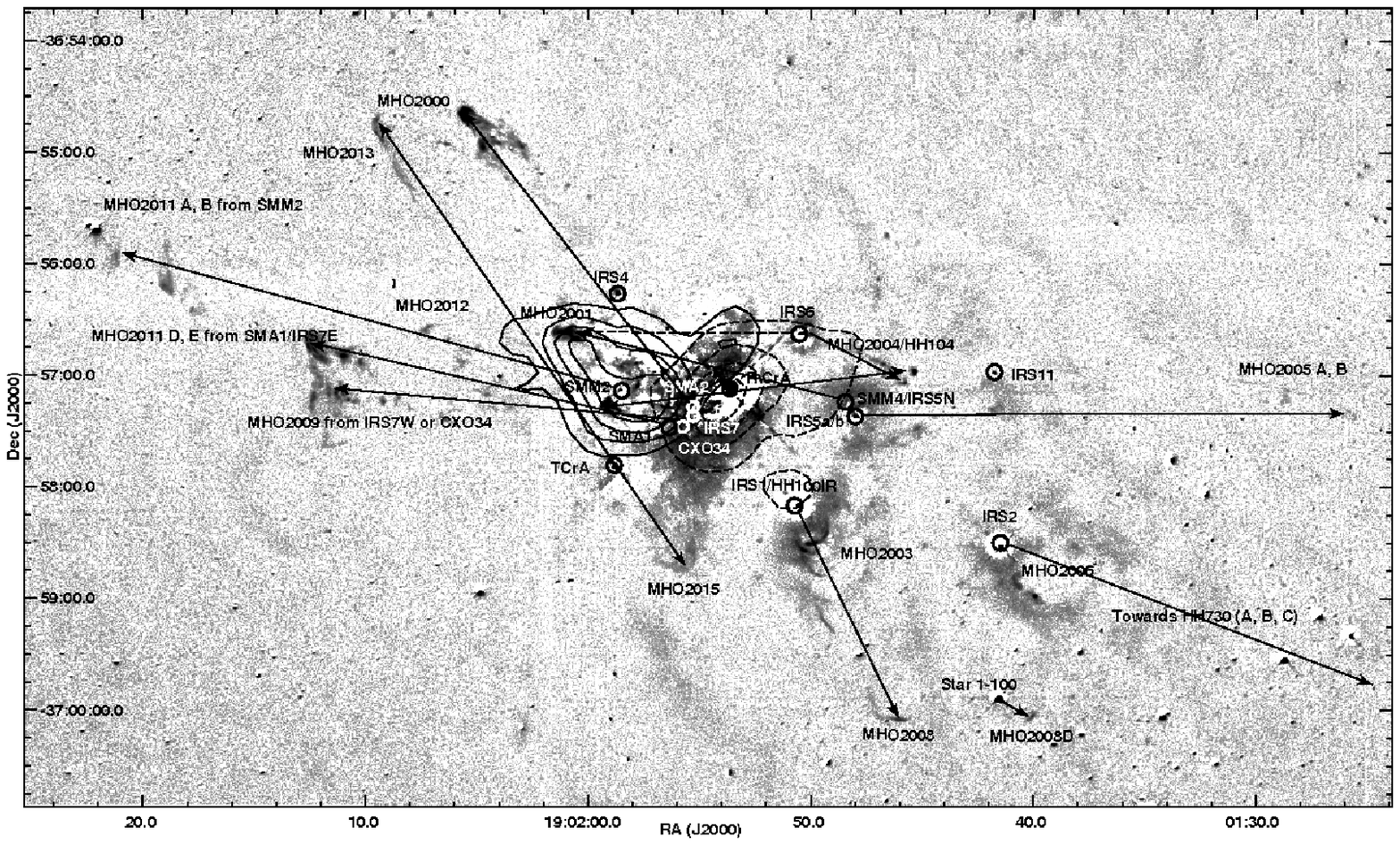}
     \caption{Continuum-subtracted H$_2$ image of the Coronet
       subregion displayed using a logarithmic scaling. The circular
       symbols show all sources detected by the {\em Spitzer} IRAC
       images. The identifications are reproduced from
       \citet{wilking97} (IRS), \citet{groppi07} (SMA), and
       \citet{nutter05} (SMM). The arrows join the probable driving
       source with the corresponding MHO. When two arrows originate
       in the same source, it implies both outflow lobes are
       visible. The CO bipolar outflow mapped by \citet{groppi04} is
       shown by solid (blue lobe) and dotted (red lobe)
       contours. For the sake of clarity, MHO 2000, which is
       situated close to SMM2, is not marked in this
       figure.}
     \label{fig:2}
\end{sidewaysfigure*}

\begin{figure*}
\centering

      \includegraphics[width=17cm,angle=0]{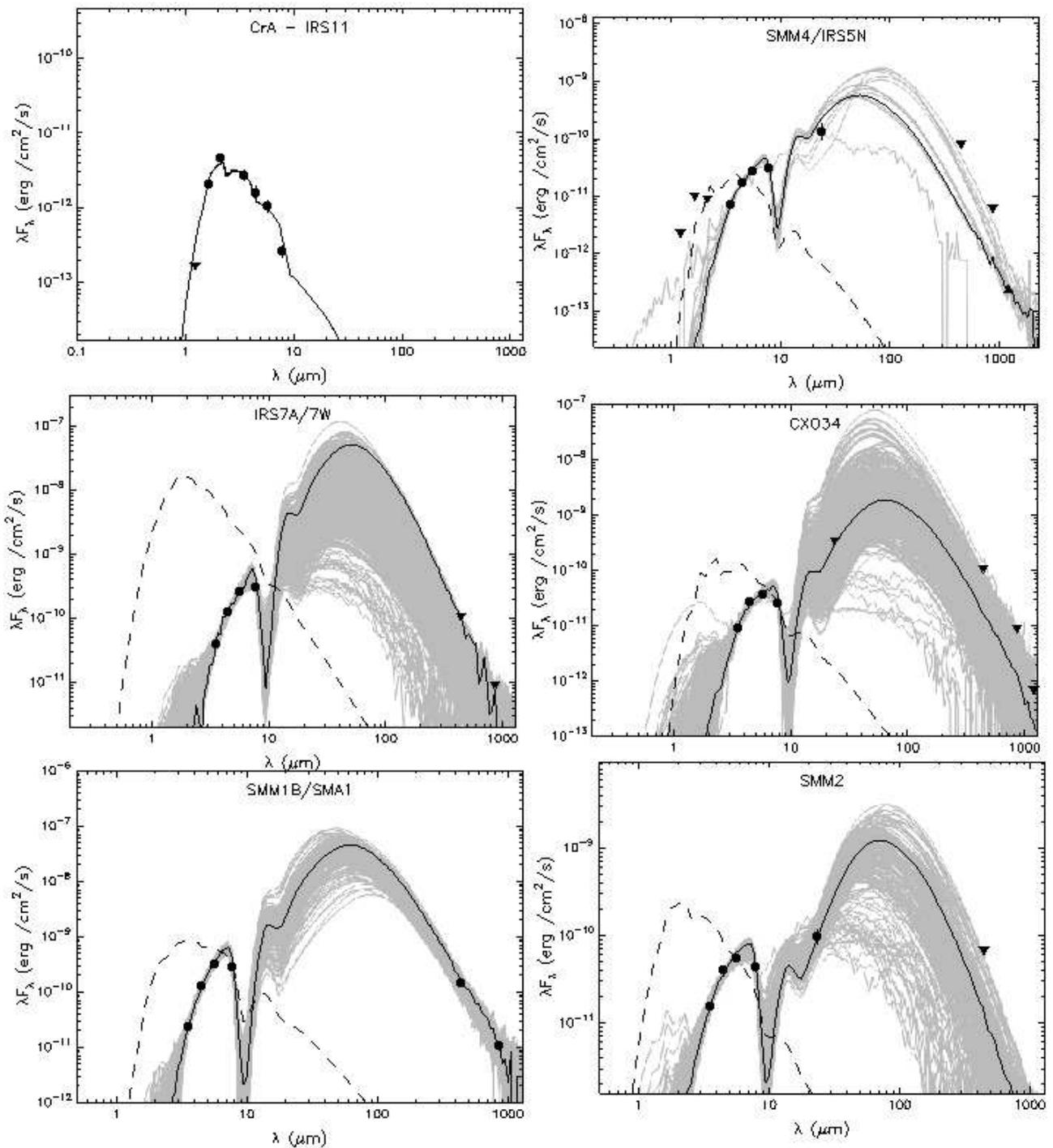}
      \caption{The observed SEDs and their best-fit models. Top left
        panel shows an example of an extincted photosphere while the
        rest are YSOs. Circular symbols represent the data
        points, upward-facing triangles are lower limits and downward
        facing triangles are upper limits. The dashed line corresponds
        to the stellar photosphere model used in the best YSO model
        fit. The gray lines show the models that satisfy the
        ${\chi}^2$ criteria described in the text. For the source
        IRS5N, the SMA data by \citet{pet11} is additionally used in
        the SED fitting.}
      \label{fig:sed_example}

\end{figure*}

\begin{table*}
 \centering
 \caption{MHO and source identifications and flow lengths}
  \label{tab:param}
\begin{tabular}{lcccccccc}

  \hline\hline\\

MHO & RA (J2000) & Dec (J2000) & Length & Source & Comments \\
\hline

2000 A, B, C & 19 02 04.8 & -36 54 45 & 198\arcsec  & SMA2 & HH99 A, B, C \\
2001 & 19 02 01.00 & -36 56 38 & 113\arcsec/145\arcsec & IRS6 or IRS5N &  \\
2002 & 19 01 59.1 & -36 57 17 & 70\arcsec & RCrA or CXO34 \\
2003,2007,2008 & 19 01 46.0 & -37 00 06 & 350\arcsec & IRS1/HH100IR & HH96, HH97, HH100, HH101\\
2004 & 19 01 46.8 & -36 56 55 & 64\arcsec & IRS6 and/or RCrA & HH104 (jet like) \\
2005 A, B & 19 01 25.9 & -36 57 20 & 258\arcsec  & IRS5a/b & thin faint arc\\
2005,C,D,E.. & 19 01 15.6 & -36 57 25 & 87\arcsec & SCrA & HH82 A, B \\
2006 & 19 01 39.2 & -36 59 02 & 330\arcsec  & IRS2 & HH730 A, B, C \\
2008 D & 19 01 40.0 & -37 00 05 & 16\arcsec  & Star1-100& 1-100HH \\
2009 & 19 02 12.2 & -36 56 43 & 193\arcsec & IRS7A & HH735, HH734 \\
2011 D,E & 19 02 12.2 & -36 56 43 & 193\arcsec & SMA1/IRS7B & HH735, HH734 \\
2011 A,B, 2012 & 19 02 20.6 & -36 55 55 & 275\arcsec & SMM2 & faint bow shocks \\
2013 & 19 02 09.2 & -36 54 53.0 & 289\arcsec & T CrA & northern lobe, HH733 (bowshock) \\
2015 & 19 01 55.7 & -36 58 42 & 67\arcsec & T CrA & southern lobe bowshock \\
2014 & 19 01 57.5 & -36 52 12 & 236\arcsec & SMM6/MMS10-11 ? & HH732 \\
2016 & 19 01 39.5 & -36 53 02 & - & TYCrA and HD176356 & - \\

\hline
\end{tabular} 
\end{table*}

\begin{table*}
 \centering
 \caption{SED-fitted parameters of selected sources in the
   Coronet. For each parameter, the minimum, maximum, and weighted mean
   (inside parantheses), selected by the ${\chi}^2$ criteria discussed
   in the text are displayed. These values represent the group of
   models shown with gray lines in Fig.\,4. SMM identifications are
   from \citet{nutter05}, IRS identifications from \citet{wilking97},
   SMA identifications are from \citet{groppi07}, and CXO
   identification is from \citet{for07}.}
  \label{tab:param}
\begin{tabular}{lcccccccc}

\hline\hline
source & \multicolumn{1}{c}{Class} & \multicolumn{1}{c}{$L$} & \multicolumn{1}{c}{M$_*$} & \multicolumn{1}{c}{$\dot{M}_{disk}$} & \multicolumn{1}{c}{M$_{disk}$} \\ 

 & & log($L_{\odot}$) & \Msun & log(\Msun $yr^{-1}$) & log(\Msun) \\
\hline

IRS5N/SMM4 & 0/I & -0.2, 1.8, (0.2) & 0.1, 2.9, (0.5) & -8.9, -6.0, (-7.3) & -3.8, -1.6, (-2.2)  \\
IRS7A/7W & 0/I & 0.1, 2.8, (1.1)  & 0.1, 6.8, (1.2) & -9.9, -4.7, (-7.0) & -4.1, -0.4, (-2.4)  \\ 
CXO34 & 0/I & -0.4, 2.5, (0.4) & 0.1, 6.5, (0.5) & -11.8, -5.2, (-8.0) & -5.0, -0.8, (-2.8) \\
SMA1 & 0/I & 1.0, 2.9, (1.9) & 0.2, 8.0, (3.3) & -9, -4.6, (-6.9) & -3.8, -0.5, (-1.7)\\ 
SMM2 & 0/I & -0.3, 2.0, (0.4) & 0.1, 4.2, (0.7) & -10.5, -6.4, (-8.1) & -4.4, -1.3, (-2.8) \\

\hline
\end{tabular} 
\end{table*}

\section{Discussion}

We discovered new MHO features tracing individual flows in the
northern part of the Corona Australis region, which encompasses the
Coronet cluster. Together with the previously known flow
  components, we have cataloged 16 individual flow lobes in the
  studied region (Table.1). Eleven of these flows are localized in the
  Coronet cluster region, which has a membership of about 14 stars
  that are detected in the x-ray \citep{for07} near-IR, mid-IR, FIR
  ({\em Spitzer}) \citep{pet11}, and submm photometric data
  \citep{nutter05,groppi07,pet11}. Thus, roughly 80\% or more of the
  embedded sources in the Coronet cluster display an active outflow
  component. The region mapped in H$_2$ here is roughly similar to the
  region covered by \citet{for07} in the x-ray and {\em Spitzer}
  data. These authors identify nine embedded objects (meaning Class
  0/I and flat-spectrum sources) in this region. Together with the
  newly identified SMA sources \citep{groppi07,pet11}, all of the
  embedded sources are associated with MHO flows.

 If we compare the submm cores from \citet{nutter05} with the detected
 MHO flows in this region, all cores except SMM1A and SMM6 are
 associated with a detected infrared source driving an outflow. SMM1A
 and SMM6 are the two cores that these authors classified as
 prestellar, of which we associate SMM6 with MHO2014, even though no
 infrared source is detected with this core up to 24$\mum$. Therefore,
 the studied region contains one, or at most two, starless cores and eight
 protostellar cores, which implies roughly 10\% of starless cores in
 the region. The Corona Australis core is therefore a good example of
 a young star-forming cluster, with a very high rate of detected
 outflow activity and young stars.

From Fig.\,3, we can see that for each driving source, only one of the
outflow lobes is traced either through H$_2$ emission or HH objects. T
CrA is the only exception in this region, associated with two lobes
identified with bow shock features. For all the other flows, the
identified flow lobes are oriented in such a way that they are flowing
in a direction away from the central dense core and entering the less
dense medium. The outflow lobes flowing towards the central dense core
is usually absent.  The dense core in the Coronet cluster is roughly
centered on the Herbig Ae/Be star RCrA \citep{chini03}.  We can see
from Fig.\,3 that the flow lobes from the sources to the left (east)
of RCrA are all moving towards the east or northeast, while their
eastern/southeastern counter-lobes are not detected. The sources
located to the right (west) of RCrA display a similar, complimentary
behavior. It appears that the outflows driven by the young stars are
``stunted'' because of the increased power needed to penetrate the
dense gas inside the core.  In comparison, the flow from TCrA, which
is orientated north-south and avoids the dense core, is bipolar.  The
dense core around Coronet is concentrated within
a 3\arcmin-4\arcmin diameter. Since the detected single lobe lengths
surpass this value, even assuming different inclinations, the
brightest bow shocks or knots from the counterlobes should have been
detected in our deep H$_2$ images, if the ``stunting'' effect was not
real. Comparing Coronet to NGC1333, we can see further support of the
likely ``stunting'' effect scenario. NGC1333 is another nearby region
with a high incidence of detected MHOs \citep[see][and references
  therein]{davis08}. In NGC1333, MHOs are associated with bipolar
lobes of multiple outflows, even in a relatively shallow survey of
\citet{davis08}. However, the dense cores are ordered along a curved
filament with gaps of low density regions \citep[see Figs.\,3 \& 7
  of][]{walsh07}, and the bipolar outflows can be seen punching out
roughly perpendicular to the filamentary axis. The MHOs detected in
this region can then be verified by comparing with Fig.5 of
\citet{davis08}. Therefore, we speculate that the higher extinction
dense core, has an important effect on the optical and infrared
detection of the flow lobes.  Only the lobes moving away to the rarer
  medium produce significant H$_2$ emission resulting in good
  detections. This observational feature may be a qualitative
  indication of the lower density of the outflowing gas compared to the
  density of the natal dense cores in which they form. It may also
  demonstrate how the outflowing gas penetrates out into a region of
  least pressure.

{\em The nature of driving sources and associated flows:} It can be
seen from Table\,2 that more than half of the detected flows, in
particular the flows originating from the Coronet, display a single
lobe length of $\ge$200\arcsec. This implies projected total flow
lengths of $\ge$0.1pc for the bipolar outflows at a distance of
130pc. By assuming roughly similar inclination angles to the observed
flows, a canonical jet speed of 100\,km\ s$^{-1}$, a flow length of
0.1pc indicates a dynamical time scale of $\sim$1200\,yr for the
outflows.  Since the molecular hydrogen flows are driven by YSOs in
their early evolutionary stage and since roughly 80\% of such
identified members are located in the Coronet cluster and are driving
H$_2$ flows, the cluster should be very young with an age
$\sim$1Myr. \citet{mw09} have evaluated the masses and ages of the
Coronet sources using near-infrared spectra and evolutionary
models. They find an age spread of 0.3--3\,Myr and masses in the range
0.2--2.5\msun for the young aggregate.

\begin{acknowledgements}
 We thank Christopher Groppi for providing the CO data for
 overplotting in Fig.\,3.  Kumar is supported by a Ci\^encia 2007
 contract, funded by FCT/MCTES (Portugal) and POPH/FSE (EC). SS and JB
 are supported by FONDECYT No.1080086, by the Ministry for the
 Economy, Development, and Tourism's Programa Inicativa Cient\'{i}fica
 Milenio through grant P07-021-F, awarded to The Milky Way Millennium
 Nucleus and from Comitee Mixto ESO-GOBIERNO DE CHILE. SS received
 partial support from Center of Excellence in Astrophysics and
 Associated Technologies BASAL CATA PFB-06. This work is based in part
 on archival data obtained with the Spitzer Space Telescope, which is
 operated by the Jet Propulsion Laboratory, California Institute of
 Technology under a contract with NASA. Support for this work was
 provided by an award issued by JPL/Caltech.

\end{acknowledgements}

\bibliographystyle{aa}

\end{document}